\documentclass[preprint,12pt,3p]{elsarticle}
\usepackage{amssymb}
\usepackage{amsmath}
\usepackage[pdftex]{color}
\usepackage{lineno}
\usepackage{hyperref}
\usepackage{cleveref}
\biboptions{sort&compress}
\usepackage[font=footnotesize,labelfont=bf]{caption}
\usepackage[font=footnotesize,labelfont=bf]{subcaption}

\begin{document}

\begin{frontmatter}

\title{\textbf{Revamped Bi-Large neutrino mixing with Gatto-Sartori-Tonin like relation }}
\author[label1]{Subhankar Roy\corref{cor1}\fnref{label3}}
\address[label1]{Department of Physics, Gauhati University, Guwahati-781014, India }
\cortext[cor1]{corresponding author}
\ead{subhankar@gauhati.ac.in, meetsubhankar@gmail.com}
\author[label5]{K. Sashikanta Singh}
\address[label5]{Department  of  Physics,  Manipur  University,  Imphal,  Manipur-795003,  India}
\ead{ksm1skynet@gmail.com}
\author[label6]{Jyotirmoi Borah}
\address[label6]{Department  of  Physics, Indian Institute of Technology Guwahati,  Guwahati-781039, India}
\ead{borah176121103@iitg.ac.in}

\begin{abstract}
The Gatto Sartori Tonin (GST) relation which connects the Cabibbo angle and the quark mass ratio: $\theta_{C}=\sqrt{m_d/m_s}$, is instituted as $\theta_{13}=\sqrt{m_1/m_3}$ to a Bi-large motivated lepton mixing framework that relies on the unification of mixing parameters:\,$\theta_{13}=\theta_{C}$ and $\theta_{12}=\theta_{23}$. This modification, in addition to ruling out the possibility of vanishing $\theta_{13}$, advocates for a nonzero lowest neutrino mass and underlines the normal ordering of the neutrino masses. The framework is further enhanced by the inclusion of a charged lepton diagonalizing matrix $U_{lL}$ with $ (\theta_{12}^l \sim \theta_C)$. The model is framed at the Grand unification theory (GUT) scale. To understand the universality of the GST relation and the Cabibbo angle, we test the observational mixing parameters at the $Z$ boson mass scale. 

\end{abstract}


\end{frontmatter}


\section{Introduction}
\label{intro}
The neutrinos are the most elusive fundamental particles available in Nature. The Standard model (SM) of particle physics fails to give a vivid picture of the same. The quest to understand the underlying first principle working behind the neutrino masses and mixing mechanism takes us beyond the SM. In this article, we emphasize on the significance of the simple unification schemes in terms of the common parameters and phenomenological relation that both the lepton and quark sectors may share.   

The SM witnesses only the left-handed flavor neutrinos and the corresponding flavor eigenstates ($\nu_{eL}$, $\nu_{\mu L}$ and $\nu_{\tau L}$) are not identical to their mass eigenstates ($\nu_{1L}$, $\nu_{2L}$ and $\nu_{3L}$). If the charged lepton Yukawa mass matrix $Y_{l}$ is diagonal, the neutrino flavor eigenstates are expressed as a linear superposition of the neutrino mass eigenstates in the following way,

\begin{equation}
\label{unu}
\nu_{\alpha L}=\sum _{i=1}^{3}(U_{\nu})_{\alpha i}\nu_{iL}, \quad(\alpha=e,\mu,\tau),
\end{equation}

where, the matrix $U_{\nu}$ is known as the Pontecorvo-Maki-Nakagawa-Sakata (PMNS) matrix\,\cite{Pontecorvo:1957cp} and it preserves the information of the Lepton mixing. The matrix $U_{\nu}$ is testable in the oscillation experiments and to parametrize $U_{\nu}$, we require three angles and six phases. Out of the six phases, three are absorbed by the redefinition of the left handed charged lepton fields\,($e_{L},\,\mu_{L},$ and $\tau_{L}$). If the original framework carries a non-diagonal charged lepton Yukawa matrix $Y_{l}$, then the $ U_{\nu}$ suffers a substantial amount of correction and the PMNS matrix is redefined as,

\begin{equation}
U=U_{lL}^{\dagger}.U_{\nu},
\end{equation}  

where, the $U_{lL}$ is the left handed unitary matrix that diagonalizes, $Y_{l}^{\dagger}.Y_{l}$. The $U$ carries six observable parameters: three neutrino mixing angles: $\theta_{12}$, $\theta_{23}$ and $\theta_{13}$ which are known as solar, atmospheric and reactor angles respectively, the Dirac-type CP violating phase\,($\delta$) and two Majorana phases\,($\psi_1$ and $\psi_2$). Following the particle data group (PDG) parametrization, the $U$ appears as shown below\,\cite{PhysRevD.98.030001},
\begin{equation}
\label{pdg}
U=R_{23}(\theta_{23}). W_{13}(\theta_{13};\delta).R_{12}(\theta_{12}).P,
\end{equation} 
where, $P= diag(e^{-i\frac{\psi_1}{2}},e^{-i\frac{\psi_2}{2}},1)$. This is to be emphasized that the oscillation experiments cannot witness the Majorana phases: $\psi_1$ and $\psi_2$ and the above parametrization ensures this fact. Moreover, the proper ordering and exact information of the neutrino mass eigenvalues are unavailable as the oscillation experiments are sensitive only to the parameters: $\Delta m_{21}^2=m_2^2-m_1^2$ and $|\Delta m_{31}^2|=|m_3^2-m_1^2|$. In short, the experimental results suggest: $\theta_{12}\approx 34^0$, $\theta_{23}\approx 47^{0}$, $\theta_{13}\approx 8^{0}$, $\Delta\,m_{21}^2=7.5\times10^{-5}\,eV^{2}$, $|\Delta\,m_{31}^2|=2.5\times10^{-3}\,eV^{2}$ and $\delta\sim 281 ^{0}$\,\cite{deSalas:2017kay}.

A specific model predicts a testable $U$. For example, the very popular mixing scheme, Tri-Bimaximal (TBM)\,\cite{Harrison:2002er} predicts the mixing angles within $U$ as, $\theta_{12}=35.26^{0}$ and $\theta_{23}=45^0$. These predictions fit well within the $3\sigma$ bound\,\cite{deSalas:2017kay} of experimental data. Hence, TBM mixing is still relevant as a first approximation. However, the former projects $\theta_{13}$ as zero and this possibility is strictly ruled out by the recent experiments\,\cite{An:2012eh,An:2016ses}.

The experiments show that
\begin{equation}
\theta_{13}\sim\mathcal{O}(\theta_{C}),
\end{equation}

where, the parameter, $\theta_{C}$ is the Cabibbo angle\,\cite{Cabibbo:1977nk} and $\theta_{C}\sim 13^{\circ}$. Hence, we expect a correction to the TBM model\,\textcolor{blue}{\cite{Xing:2002sw}} which is of the order of $\theta_{C}$. Another example of mixing scheme that carries vanishing $\theta_{13}$ is the democratic mixing pattern which predicts large solar and atmospheric angles: $\theta_{12}=45^{\circ}$ and $\theta_{23}\simeq 54.7^{\circ}$. But interestingly, in order to converge to the reality conditions, all these three mixing angles require corrections of the order of $\theta_{C}$. In Ref.\,\cite{Xing:2011at}, the natural perturbation is implemented on the democratic mixing matrix, $U_{0}$ and the PMNS matrix is defined as, $U=U_{0}X$, where the $X$ is the correction matrix such that: $X=X\,(\theta_{13}^{x},\theta_{12}^{x},\theta_{23}^{x} )$. From, the model-building point of view, on choosing $\theta_{13}^{x}=0$ and sticking to the \textit{ansatz}: $\theta_{12}^{x} \simeq - \theta_{23}^{x} \simeq \theta_{C}$, one may see that except $\theta_{23}$, the other two angles remain slightly outside the $3\sigma$ bounds\,\cite{deSalas:2017kay}. But whether $U_{0}$ arises from the neutrino sector or charged lepton sector (or both) is model-dependent. 

On the other hand, the promising mixing schemes termed as Bi-large(BL) neutrino  mixing\,\cite{Boucenna:2012xb, Ding:2012wh, Branco:2014zza,Roy:2012ib, Roy:2014nua, Ding:2019vvi, Chen:2019egu} shelters $\theta_{C}$ as an inherent parameter within the neutrino sector. Also, it assumes, large\,(and equal) values of $\theta_{12}$ and $\theta_{23}$. The angle, $\theta_{13}$ is visualized as: $\sin\theta_{13}\sim \lambda$, where $\lambda=\sin\theta_{C}\approx 0.22$, is called the Wolfenstein parameter\,\cite{Wolfenstein:1983yz}. The geometrical origin of the BL model is explored originally in Ref.\,\cite{Ding:2012wh}. We know that $\theta_{C}$ is a significant parameter within the quark sector and realization of the same within the neutrino sector extends the possibilities for new unification schemes. The BL framework is further strengthened by the fact that in the $SO(10)$ or $SU(5)$ inspired GUTs, a single operator generates the Yukawa matrices for both: down type quarks and the charged leptons\,($Y_{d}$ and $Y_{l}$ respectively)\,\cite{Pati:1974yy,Elias:1975kf,Elias:1977bv, Blazek:2003wz,Dent:2007eu,kounnas1985grand, ross2003grand}. In this context, the matrix elements of $Y_{l}$ are proportional to those of  $Y_{d}$ and this results in,

\begin{equation}
U_{lL}\sim V_{CKM}
\end{equation}

where, the $V_{CKM}$ is called the Cabibbo-\,Kobayashi-\,Maskawa matrix\,\cite{10.1143/PTP.49.652, Wolfenstein:1983yz} and the PMNS matrix is redefined as $U\sim V_{CKM}^{\dagger}.U_{\nu}$ \cite{Duarah:2012bd} in a basis where $Y_{l}$ is diagonal. 
Interestingly, the role of the Cabibbo angle is not limited to the quark mixing only, but it describes the quark masses also. We see that, ratio of up and charm quark masses: $m_{u}/m_{c}\sim \lambda^4$ and that between charm and top quarks: $m_c/m_t\sim \lambda^4$. Also, the ratio of down and strange quarks and that between strange and bottom quarks are $m_{d}/m_s\sim\lambda^2$ and $m_{s}/m_b \sim \lambda^2$ respectively\,\cite{XING20201}. The former case is called the Gatto-Sartori-Tonin (GST) relation and is expressed as shown below\,\cite{Gatto:1968ss}:

\begin{equation}
\label{gst}
\sin\theta_{C}\simeq\sqrt{\frac{m_d}{m_s}},
\end{equation} 

The above relation is derived in many occasions starting from the study of discrete flavor symmetry groups\,\cite{doi:10.1111/j.2164-0947.1977.tb02958.x, WILCZEK1977418, FRITZSCH1977436, PAKVASA197861}. The question appears  whether in case of lepton sector, the masses and the mixing angles are somehow related or not. Indeed, the quark and the neutrino sector differ a lot than being similar. The $V_{CKM}$ is too close to an Identity matrix, whereas the PMNS matrix $U$, is far from being an Identity matrix. Although the mixing schemes differ a lot, but believing on the unification framework like GUT, there lies enough reasons to explore similar signatures in both quark and lepton sectors. We see that in the lepton sector the ratios of the charged lepton masses: $m_e/m_{\mu}\sim \lambda^2$ and $m_{\mu}/m_{\tau}\sim \lambda^2$\,\cite{XING20201}. So, we see that even though there are differences, yet the parameter $\theta_{C}$\,(or $\lambda$) finds its existence in both quark and lepton sectors. But unlike the charged leptons, the exact masses of the neutrino mass eigenstates are not yet known.

Following the footprints of GST relation in eq.(\ref{gst}), the viability of a similar GST like relation in the neutrino sector:
\begin{equation}
\label{ngst}
\sin\theta_{ij}=\sqrt{\frac{m_i}{m_j}},
\end{equation}   
is explored in our earlier work\,\cite{Roy:2015cza}. In this analysis, the $Y_{l}$ is assumed to be diagonal and the CP violation is ignored. The phenomenology shows that there is a single GST like relation,

\begin{equation}
\sin\theta_{13}=\sqrt{\frac{m_1}{m_3}},
\end{equation}
or,
\begin{equation}
\sin\theta_{23}=\sqrt{\frac{m_3}{m_2}},
\end{equation}
which is possible in the neutrino sector. Needless to mention that the two relations cannot be experienced simultaneously. The first relation seems more appealing when we are trying to explore the unification possibilities. This is because $\theta_{13}$ and $\theta_{C}$ are of the same order and in this work we assume that $\theta_{13}$ unifies with $\theta_{C}$ at the GUT scale.
Also, this relation advocates for the normal ordering of the neutrino masses and this possibility is indicated recently by the experimental results\,\cite{deSalas:2017kay}. On the other hand, the second relation concerning $\theta_{23}$ favors inverted ordering of neutrino masses. The vindication of nonzero $\theta_{13}$, its proximity towards the Cabibbo angle and the hint for normal ordering of neutrino masses make the foundation of unification schemes stronger. In the next section we shall try to explore how the GST relation can be invoked in the framework of Bi-large neutrino mixing.

\section{Modified bilarge ansatz}


Several BL schemes are proposed in the Refs.\,\cite{Boucenna:2012xb, Ding:2012wh, Branco:2014zza, Roy:2012ib, Roy:2014nua, Ding:2019vvi, Chen:2019egu}, and out of which we adopt the one\,\cite{Boucenna:2012xb,Roy:2014nua} which in addition to unifying the reactor angle and Cabibbo angle, $\theta^{\nu}_{13}=\theta_{C}$ stresses further on the unification of the atmospheric and solar mixing angles such that within the neutrino sector, $\sin\theta^{\nu}_{12}=\sin\theta^{\nu}_{23}=\psi \lambda$. Here, $\psi$ is a free parameter and as per the earlier analysis\,\cite{Boucenna:2012xb, Roy:2012ib, Roy:2014nua}, $\psi\approx 3$. In our previous works\,\cite{Roy:2012ib, Roy:2014nua}, it is shown that such a BL mixing framework can be made more promising by incorporating a CKM-like charged lepton diagonalizing matrix. But none of the works related to BL mixing mentioned above takes the neutrino mass parameters into consideration. Emphasizing on the possibility that there may lie a correlation between the reactor angle and the mass ratio, $m_1/m_3$, and this relation exists naturally in a unification framework defined at the GUT scale\,($\sim 10^{16}$ GeV), the Bi-large \textit{ansatz} in the lepton sector is modified as presented below.

\begin{eqnarray}
\theta^{\nu}_{13}&=&\theta_{C}=\sqrt{\frac{m_{1}}{m_{3}}}=\sqrt{\frac{m_{d}}{m_{s}}},\\
\theta^{\nu}_{12}&=&\theta^{\nu}_{23}=\sin^{-1}(\psi \lambda),\\
\theta_{12}^{l}&\simeq & \theta_{C}, \\
 \theta^{l}_{23}&=& A\lambda^2,
\end{eqnarray}

where, the $\theta_{ij}^{\nu}$'s and $\theta_{ij}^{l}$'s are the mixing angles that parametrize the $U_{\nu}$ and $U_{lL}$ respectively. Here, $A$ is one of the Wolfenstein parameters that appears in the CKM matrix\,\cite{Charles:2004jd}. It is worth mentioning that this modified BL framework favors the normal ordering of the neutrino masses. At the same time, we see that the provision of \textit{strict} normal hierarchy which insists on $m_1=0$ is ruled out as $\theta_{13}^{\nu}$ is nonzero. The last two \textit{ansatze} involving the mixing angles from charged lepton sector indicates for a CKM-like $U_{lL}$.

Based on the above discussion, at the GUT scale we design the neutrino mixing matrix as shown below,
\begin{eqnarray}
U_{\nu}=\left(\begin{array}{ccc}
 c-\frac{c \lambda ^2}{2} & s-\frac{s \lambda ^2}{2} & e^{-i \delta_0 } \lambda  \\
 -c s \left(e^{i \delta_0 } \lambda +1\right) & c^2-e^{i \delta_0 } s^2 \lambda  & s-\frac{s \lambda ^2}{2} \\
 s^2-c^2 e^{i \delta_0 } \lambda  & -c s \left(e^{i \delta_0 } \lambda +1\right) & c-\frac{c \lambda ^2}{2} \\
\end{array}\right).P, 
\end{eqnarray}

where, $s=\psi\lambda$ and $s=\cos(\sin^{-1}(\psi\lambda))$. The $\delta_{0}$ is a free phase parameter within the neutrino sector at the GUT scale. We establish the effective light neutrino mass matrix, $m_{\nu}$ as shown below,

\begin{eqnarray}
\label{mnu}
m_{\nu}(m_2,m_3,\psi,\psi_1,\psi_2,\delta_0,\lambda)= U_{\nu}^{*}.diag\lbrace \lambda^2, m'_{2},1 \rbrace. U_{\nu}^{\dagger}\, m_{3},
\end{eqnarray}
where, $m'_2=m_2/m_3$. The $m_{\nu}$  contains four free parameters: $m_3$, $m_2$, $\psi$, $\delta_0$ and two Majorana phases $\psi_1$ and $\psi_2$. The $m_\nu$ depends on the renormalization energy scale $\mu$. As $m_\nu$ appearing in the equation above is defined at $M_{GUT}$, we have to run it down upto testable low energy scale, at $M_{Z}$, the Z boson mass scale to extract the information of the oscillation parameters from the former. The running of $m_\nu$ involves several complicated steps. We believe that $m_\nu$ results from the see saw mechanism\,\cite{Minkowski:1977sc, Mohapatra:1979ia}  which again involves three heavy right-handed Majorana neutrino mass eigen states, $N_{R}^{i=1,2,3}$ with the mass eigenvalues $M_{R}^{1,2,3}$ respectively. Here, the $N_{R}^3$ is the heaviest Right-handed eigenstate and $M_{R}^3< M_{GUT}$. The effective light neutrino mass matrix, $m_{\nu}(\mu)$ is related to heavy right handed neutrino mass matrix, $M_{R}(\mu)$ and light Dirac neutrino Yukawa matrix, $Y_{\nu}(\mu)$ in the following way,

\begin{equation}
m_{\nu}(\mu)=-\frac{v^2}{2} Y_{\nu}(\mu)^T M_{R}^{-1}(\mu)Y_{\nu}(\mu),
\end{equation}

where, $v$ is the Higgs vev. In our analysis, we assume that this parameter does not run. We shall be working in the light of minimal super symmetric extension of the SM (MSSM)\,\cite{Randall:1998uk, Haber:1990aw, Dine:1995ag} and thus we take, $v=246\,GeV\,\sin\beta$.  While running down the $m_{\nu}$, the heavy right-handed states are to be integrated out at different thresholds. 

Between $M_{GUT}$ and $M_{R}^{3}$, the following Renormalization Group Equation\,(RGE) holds good\,\cite{Balaji:2000au, Mohapatra:2003tw, Antusch:2002fr, Joshipura:2002xa, Antusch:2003kp, Ohlsson:2013xva, Gupta:2014lwa, Haba:2014uza, Hollik:2014hya, Singh:2015kua, Singh:2018cxy}.

\begin{eqnarray}
16\pi^2 \frac{d m_{\nu}}{dt}&=&\left(-\frac{6}{5} g_{1}^2-6g_{2}^2 + 2 Tr (Y_{\nu}^{\dagger}Y_{\nu}+3\,Y_{u}^{\dagger}.Y_{u})\right)m_{\nu}\nonumber\\
 &&+(Y_{l}^{\dagger}Y_{l}+ Y_{\nu}^{\dagger}Y_{\nu})^{T}m_{\nu} +m_{\nu}(c_{l}Y_{l}^{\dagger}Y_{l}+c_{\nu}Y_{\nu}^{\dagger}Y_{\nu}),
\end{eqnarray}

where, $t=ln(\mu/\mu_0)$. Here, $Y_{u}$ is the up quark yukawa matrix. Apart from the knowledge of $m_\nu$, $Y_{\nu}$, and $Y_{u}$ at the GUT scale, we require the information of Yukawa matrix of down quark\,($Y_{d}$) as the texture of the $Y_{l}$ is dependent on how we parametrize $Y_{d}$. To define the textures of the respective Yukawa matrices, we shall draw the motivation from SU(5) GUT phenomenology discussed in Refs.\,\cite{Antusch:2009gu, Marzocca:2011dh, Antusch:2013ti, Antusch:2013rxa, Roy:2015cka,Rahat:2018sgs}. 

First, we parametrize $Y_{d}$ as shown below,

\begin{eqnarray}
\label{yd}
Y_{d} &=& 
\left(\begin{array}{ccc}
 0 & d_{12}\lambda^5 & 0 \\
 d_{21}\lambda^5 & d_{22}\lambda^4 & -d_{23}\lambda^3 \\
 d_{31}\lambda^7 & d_{32}\lambda^6 & d_{33}\lambda  \\
\end{array}\right),
\end{eqnarray}
where, the $d_{ij}'$\,s are $\mathcal{O}(1)$ coefficients and these are tabulated in Table.\,(\ref{Table1}). 

The SU(5) GUT suggests that a general element of the charged lepton Yukawa matrix,$(Y_{l})_{ij}$ is proportional to $(Y^{T}_{d})_{ij}$, $(Y_{l})_{ij} = \alpha\,(Y_{d})_{ji}$. Here, the proportionality constant $\alpha$ is not arbitrary, rather it's choice is strictly guided by SU(5) GUT phenomenology and is constrained to limited numbers of integers and fractions described in the Refs.\,\cite{Antusch:2009gu, Marzocca:2011dh, Antusch:2013ti, Antusch:2013rxa}. We choose $\alpha$ in terms of allowed entries: $\lbrace -\frac{2}{3},\, \frac{1}{2},\,\frac{3}{2},\,6\rbrace$ and portray the charged lepton Yukawa matrix as shown below,
\begin{eqnarray}
Y_{l} &=& 
\left(\begin{array}{ccc}
0 & 6\, d_{12}\lambda^5 & 0 \\
-\frac{1}{2}\, d_{21}\lambda^5 & 6\, d_{22}\lambda^4 & \frac{3}{2}\,d_{23}\lambda^3 \\
-\frac{1}{2}\, d_{31}\lambda^7 & 6\, d_{32}\lambda^6 & -\frac{3}{2}\,d_{33}\lambda  \\
\end{array}\right)^T.
\end{eqnarray}
In developing the above two yukawa matrices, we take care of the ratio $(y_{\mu}y_{d})/(y_{s} y_e)$ which comes out to be 11.40 and this complies with the bound prescribed by the Ref.\, \cite{Antusch:2013rxa}. We choose the non-diagonal up-quark yukawa matrix, $Y_{u}$ in the following manner,

\begin{equation}
\label{yu}
Y_{u}=\left(\begin{array}{ccc}
 u_{11}\lambda^8 & 0 & 0 \\
 0 & u_{22}\lambda^4 & -u_{23}\lambda^6 \\
 0 & u_{32}\lambda^3 & u_{33}  \\
\end{array}\right),
\end{equation}  
where, $u_{ij}$s are $\mathcal{O}(1)$ coefficients\,(see Table.\,(\ref{Table1})). We diagonalize  the above matrices following the \textbf{RL} convention such that, $U_{(x)R}^{\dagger}.Y_{(x)}. U_{(x)L}=Y_{(x)}^{diag}$, where, $x=d,u$ and $l$. The matrices $U_{(x)L}$ are recognized by diagonalizing $Y^{\dagger}_{(x)}Y_{(x)}$ such that $U_{(x)L}^{\dagger}.Y^{\dagger}_{(x)}Y_{(x)}. U_{(x)L}=Y_{(x)}^{diag^2}$. We identify,  
\begin{eqnarray}
U_{uL}&\approx &\left(\begin{array}{ccc}
1 & 0 & 0\\
0 & 1 & -A\,\lambda^2\\
0 & A\,\lambda^2 & 1
\end{array}\right),\\
U_{dL} &\approx & \left(\begin{array}{ccc}
1-\frac{\lambda^2}{2} &\lambda  & 0\\
-\lambda & 1-\frac{\lambda^2}{2} & 0\\
0 & 0 & 1
\end{array}\right),
\end{eqnarray}

and,
\begin{eqnarray}
\label{UlL}
U_{lL} &\approx & \left(\begin{array}{ccc}
1-\frac{a^2\lambda^2}{2} & a\lambda  & 0\\
-a\lambda & 1-\frac{a^2\lambda^2}{2} & -A\,\lambda^2 \\
0 & A\,\lambda^2 & 1
\end{array}\right),
\end{eqnarray}
for $Y_{u}$, $Y_d$ and $Y_l$ respectively. Here $a=1.03$ and the $V_{CKM}$ matrix is identified as, 
\begin{eqnarray}
V_{CKM} = U_{uL}^{\dagger}.U_{dL}&\approx &  \left(\begin{array}{ccc}
1-\frac{\lambda^2}{2} &\lambda  &0\\
-\lambda & 1-\frac{\lambda^2}{2} &  A\,\lambda^2\\
0 & - A\,\lambda^2  & 1
\end{array}\right)
\end{eqnarray}

We see that the parameter $a$ appearing in $U_{lL}$ shifts a little from unity and thus the $U_{lL}$ is not exactly equal to the $V_{CKM}$ matrix, but $U_{lL}\approx V_{CKM}$. The parameter $a=1$, is true if the correlation between $Y_{l}$ and $Y_{d}$ were, $Y_{l}=Y_{d}^{T}$. The choice of the Dirac neutrino Yukawa matrix, $Y_{\nu}$ is arbitrary and we fix it as per Ref.\cite{Antusch:2005gp} as shown below,

\begin{eqnarray}
\label{ynu}
Y_{\nu}= \frac{1}{2}\left(\begin{array}{ccc}
\nu_{11} \lambda^3 & 0 & 0 \\
\nu_{21}\lambda^6 & \nu_{22}\lambda & 0\\
0 & 0 & 1
\end{array}\right),
\end{eqnarray} 
where, the coefficients $\nu_{ij}$'s are illustrated in Table.\,(\ref{Table1}).

The $m_{\nu}$ suffers further Quantum corrections in the intervals, $M_{R}^2<\mu<M_{R}^{3}$ and  $M_{R}^1<\mu<M_{R}^{2}$ \,\cite{Tanimoto:1995bf, Mei:2004rn, Ohlsson:2019sja, Coy:2018bxr, Antusch:2018gnu, Huan:2018lzd, Antusch:2017ano}. In order to deal with the RGE evolution of the effective neutrino mass matrix, the Yukawa matrices, gauge couplings, to integrate out the heavy neutrino singlets and to derive the information of the neutrino oscillation parameters at different energy scales, we use the mathematica package \textbf{REAP}\,(Renormalisation group Evoluion of Angles and Phases)\,\cite{Antusch:2005gp}.
The analysis involves a parameter known as supersymmetry breaking scale ($m_{susy}$) which is still unknown. Theoretically, $m_{s}$ ranges from a few Tev to hundred Tev.

\section{Numerical Analysis}
To exemplify, we choose $M_{GUT}=4.577\times 10^{16}\,GeV$. At this scale we set, $\lambda=0.2250$, $A=0.705$ and the free parameters as shown below,\\
\begin{eqnarray}
\label{inputs}
&\psi = 2.94,\, m_2 = 0.0095\,eV,\, m_3 = 0.054\,eV,&\nonumber\\
&\delta_0= 318^{\circ},\, \psi_1 = \psi_2 = 360^{\circ} &\nonumber\\
& g_1=0.7063,\, g_2 = 0.7065,\,g_3 = 0.7069, &\nonumber
\end{eqnarray}
and fix $\tan\beta$ at $60$. We choose $m_{susy}=3$ \textit{TeV} as one out of many possibilities. 
The RGEs are run from $M_{GUT}$ upto $M_{Z}=91.1876\,GeV$ and we extract the necessary information of the observable neutrino mixing parameters. We obtain,
\begin{eqnarray}
&\theta_{12}= 34.26^{\circ},\,\theta_{13}=8.80^{\circ},\,\theta_{23}= 48.56^{\circ},&\nonumber\\
&\delta = 274.73^{\circ},\,\Delta m_{sol}^2 = 7.58 \times 10^{-5}\, eV^2, &\nonumber\\
&\Delta m_{atm}^2 = 2.51 \times 10^{-3}\,eV^2,\, \sum m_{\nu_{i}}= 0.063\,eV, &\nonumber\\
& \psi_{1}= 359.84 ^{\circ}, \, \psi_{2}= 4.04^{\circ}&\nonumber
\end{eqnarray} 

We see that the two angles $\theta_{12}$ and $\theta_{23}$ comply well within the $1\sigma$ bound, $\theta_{13}$ is consistent within the $2\sigma$ and the solar and the atmospheric mass squared differences agree to stay within the $1\sigma$ bound\,\cite{deSalas:2017kay,Esteban:2018azc}. According to the recent analysis in Refs.\,\cite{Giusarma:2016phn, Loureiro:2018pdz}, the observational parameter, $\sum m_{\nu_{i}}$ has got an upper bound of $0.15\,eV$ to $0.27\,eV$ and the most stringent upper bound is $0.078\, eV$ as per Ref.\,\cite{Choudhury:2018byy}. The lower bound is predicted as $ 0.058\,eV$ in Refs.\, \cite{Loureiro:2018pdz, Choudhury:2018byy} or $0.060\,eV$ according to the Ref.\,\cite{PhysRevD.98.030001}. We see that prediction of $\sum m_{\nu_{i}}$ in our analysis lies slightly above the prescribed lower bound.

We see that the  $\theta_{13}$ at $M_Z$, unlike the other mixing angles, varies appreciably if the free parameter $\delta_{0}$ varies at the GUT scale. To illustrate, keeping all the input parameters fixed, if $\delta_{0}$ is changed a little from $318^{\circ}$ to $323^{\circ}$, we see that the $\theta_{13}$ at the $M_{Z}$ scale changes from $8.67^{\circ}$ to $7.98^{\circ}$ (which lies outside the $3\sigma$ range).

Similarly, if $m_{susy}$ is varied a little, the mass parameters at $M_{Z}$ are also affected. To illustrate, we study how the different observational parameters at the $M_{Z}$ scale evolve against the variation of $\delta_{0}$, for different values of $m_{s}$ ranging from $1\,Tev$ to $14\,TeV$.  The analysis requires the knowledge of numerical values of the three gauge coupling constants and three Yukawa couplings at the GUT scale. For this, we use the Refs \,\cite{Singh:2015kua, Singh:2018cxy}, where the required input parameters are obtained by running the RGE s following a  bottom-up approach at different values of $m_{susy}$,\,(See Table.\,(\ref{Table2})).

\begin{figure*}
\begin{center}
\begin{subfigure}{0.48\textwidth}
\includegraphics[scale=0.8]{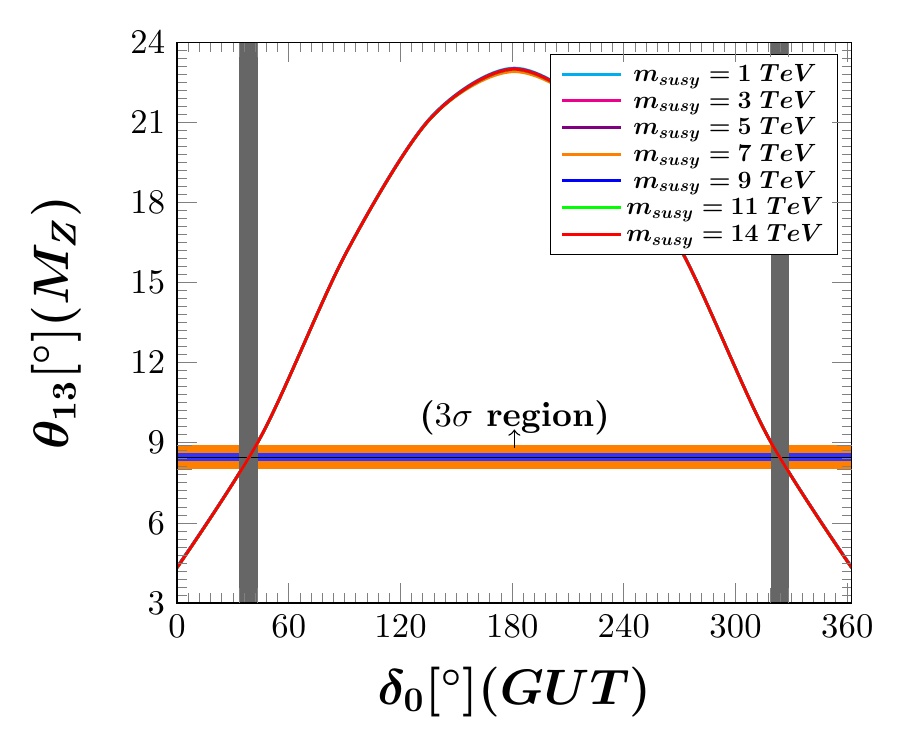} 
\caption{\label{t13}}
\end{subfigure}
\vspace*{\fill}
\begin{subfigure}{0.48\textwidth}
\includegraphics[scale=0.8]{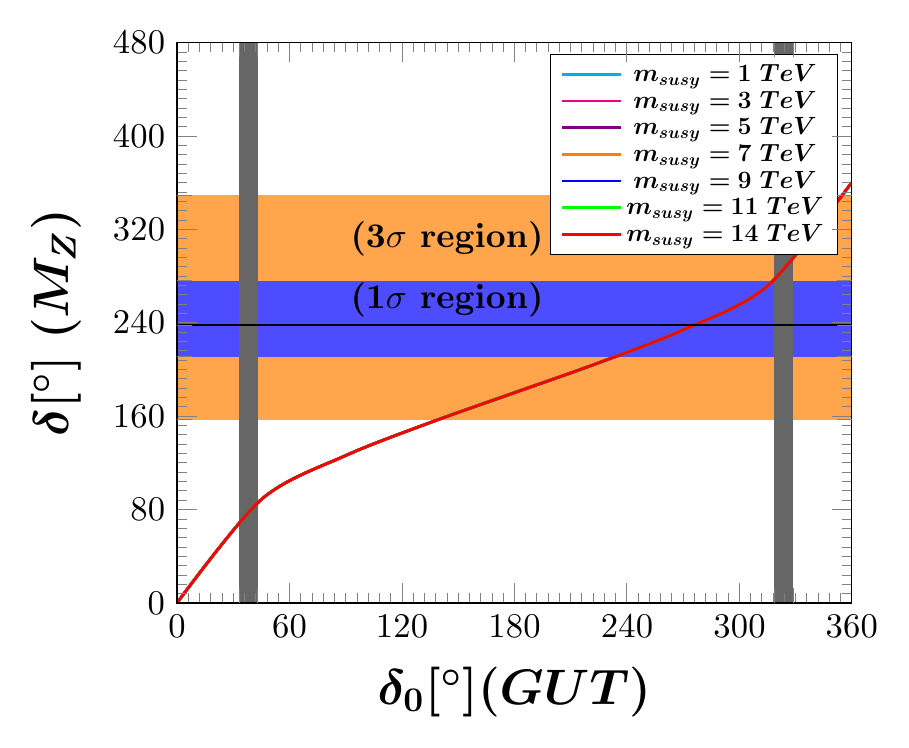} 
\caption{\label{dir}}
\end{subfigure}
\end{center}
\caption{\footnotesize (\textbf{\ref{t13})} and  \textbf{(\ref{dir})} show how the $\theta_{13}(M_{Z})$ changes with respect to variation of $\delta_{0}$ at the GUT scale respectively for different values of the SUSY breaking scale,\,$m_{susy}$ ranging from $1\,TeV$ to $14\,TeV$ (All the graphs are merged almost together). In both of the plots, black horizontal line, purple and orange bands signify the best fit value, $1\sigma$ and $3\sigma$ ranges of the concerned parameter. In Fig.\,(\textbf{\ref{t13}}), with respect to the $3\sigma$ range\,\cite{deSalas:2017kay} of $\theta_{13}$, two possible ranges of input parameter $\delta_{0}$: $33.40^{\circ}\leqslant\delta_{0}\leqslant 43.30^{\circ}$ and $318.90^{\circ}\leqslant\delta_{0} \leqslant 328.70^{\circ}$\,(shown by two vertical grey bands) are obtained. In Fig.\,(\textbf{\ref{dir}}) we see that only the second range is allowed in the light of the $3\sigma$ bound of $\delta$. This range $318.90^{\circ}\leqslant\delta_{0} \leqslant 328.70^{\circ}$  predicts the Dirac CP phase\,($\delta$) within the $2\sigma$ bound\,\cite{deSalas:2017kay}.}
\end{figure*}

\begin{figure*}
\begin{center}
\begin{subfigure}{0.48\textwidth}
\includegraphics[scale=0.8]{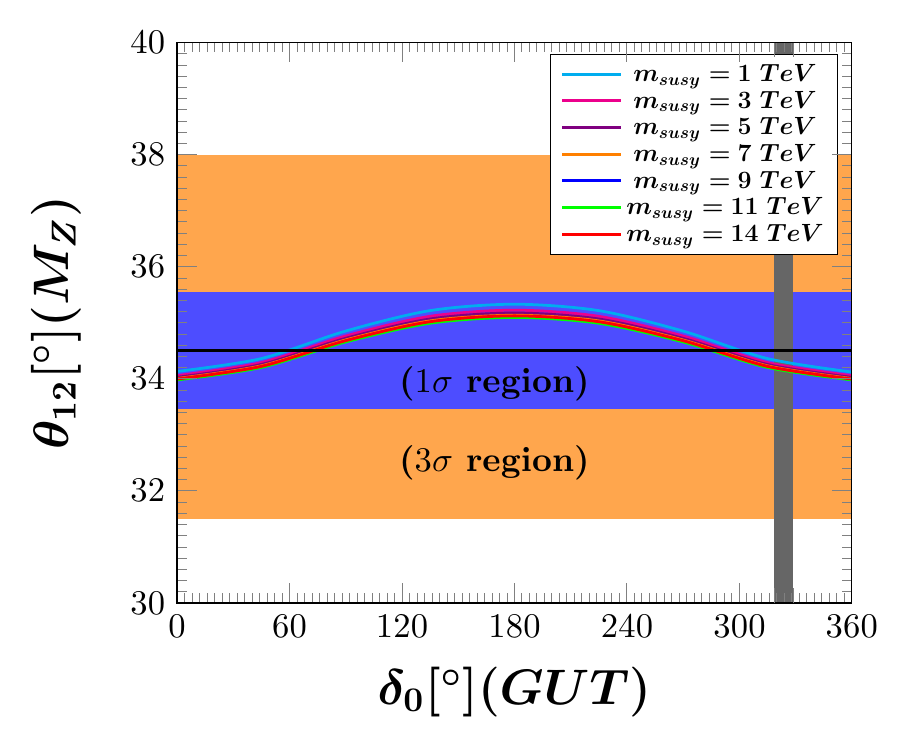} 
\caption{\label{t12}}
\end{subfigure}
\vspace*{\fill}
\begin{subfigure}{0.48\textwidth}
\includegraphics[scale=0.8]{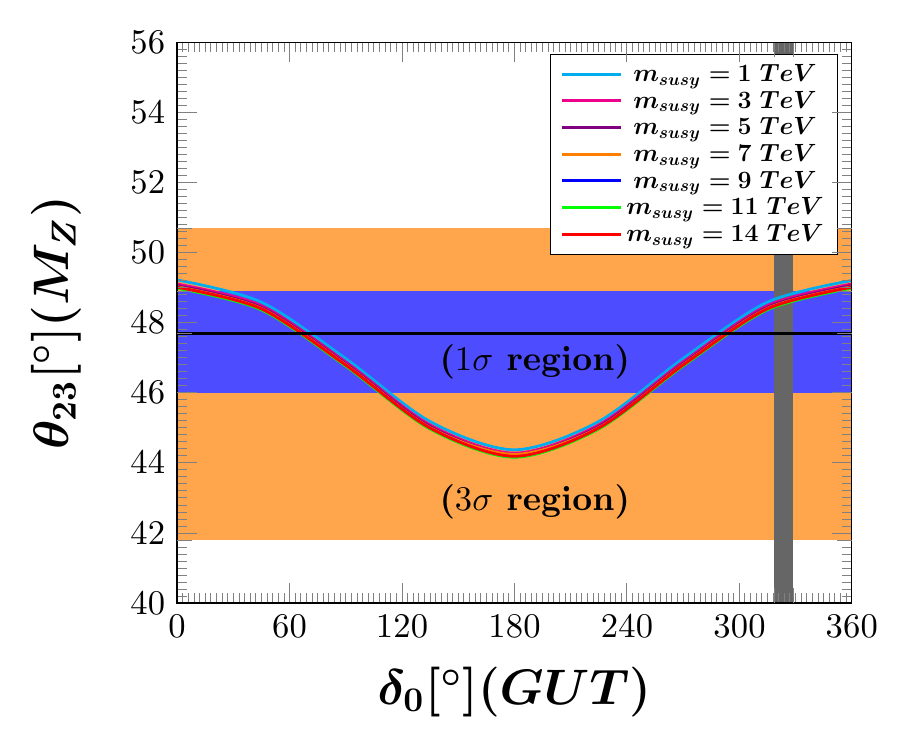} 
\caption{\label{t23}}
\end{subfigure}\\
\begin{subfigure}{0.48\textwidth}
\includegraphics[scale=0.8]{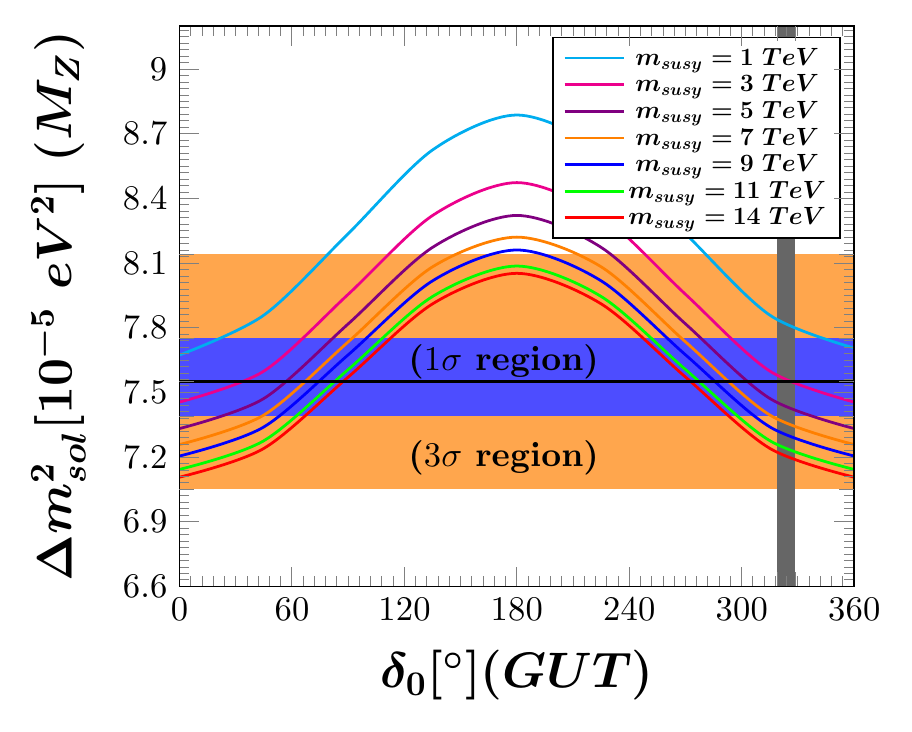} 
\caption{\label{msol}}
\end{subfigure}
\vspace*{\fill}
\begin{subfigure}{0.48\textwidth}
\includegraphics[scale=0.8]{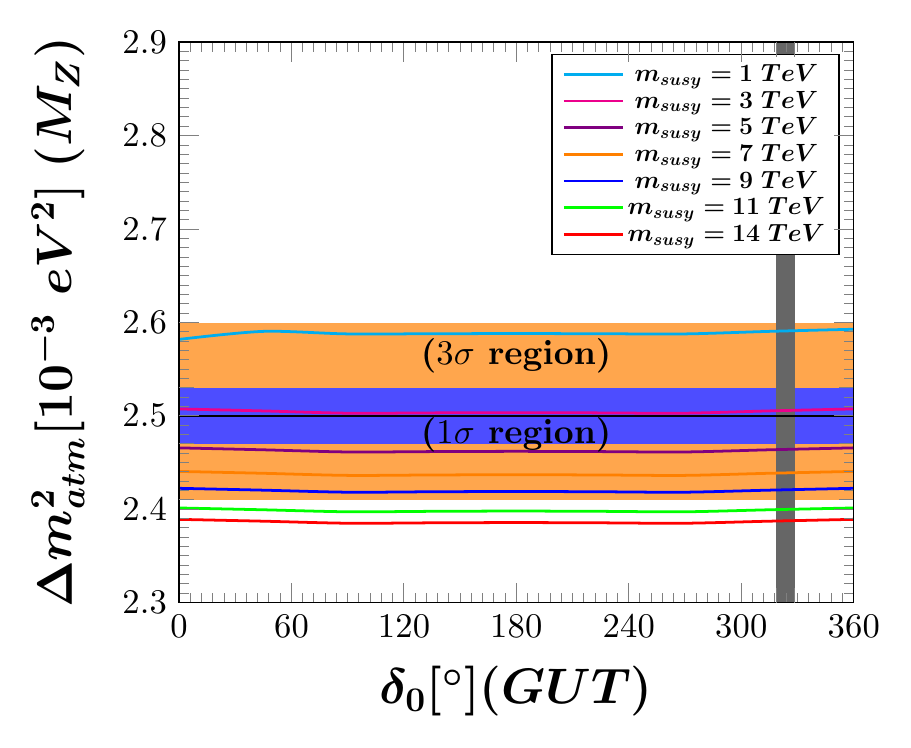} 
\caption{\label{matm}}
\end{subfigure}
\end{center}
\caption{\footnotesize \textbf{(\ref{t12})}, \textbf{(\ref{t23})}, \textbf{(\ref{msol})} and \textbf{(\ref{matm})} show the variation of $\theta_{12}$, $\theta_{23}$, $\Delta\,m_{sol}^2$ and $\Delta\,m_{atm}^2$ at the $M_{Z}$ scale respectively with respect to the variation of $\delta_{0}$ at the GUT scale for different values of the SUSY breaking scale,\,$m_{susy}$ ranging from $1\,TeV$ to $14\,TeV$. The plots in Figs\,(\textbf{\ref{t12}}) and (\textbf{\ref{t23}}) merge almost together. The Black line, purple and the orange bands represent the best-fit, $1\sigma$ and $3\sigma$ bounds \cite{deSalas:2017kay} respectively for the concerned observational parameters. The vertical grey band represents the allowed bound of $\delta_{0}$ at GUT scale which is $318.90^{\circ}\leqslant\delta_{0} \leqslant 328.70^{\circ}$. The $\theta_{12}$ is predicted around $34^{\circ}\,(1\sigma)$ and that for $\theta_{23}$ is around $49^{\circ}$\,\cite{deSalas:2017kay}.In Figs.\,(\ref{msol}) and (\ref{matm}), the $\Delta m_{sol}^2$ and $\Delta m_{atm}^2$ varies appreciably with respect to both $\delta_{0}$ and $m_{susy}$. We note that unlike $\Delta m_{sol}^2$, the $\Delta_{atm}^2$ goes outside the $3\sigma$ range if $m_{susy}> 9\,Tev$.}
\end{figure*}

\begin{figure*}
\begin{center}
\begin{subfigure}{0.48\textwidth}
\includegraphics[scale=0.8]{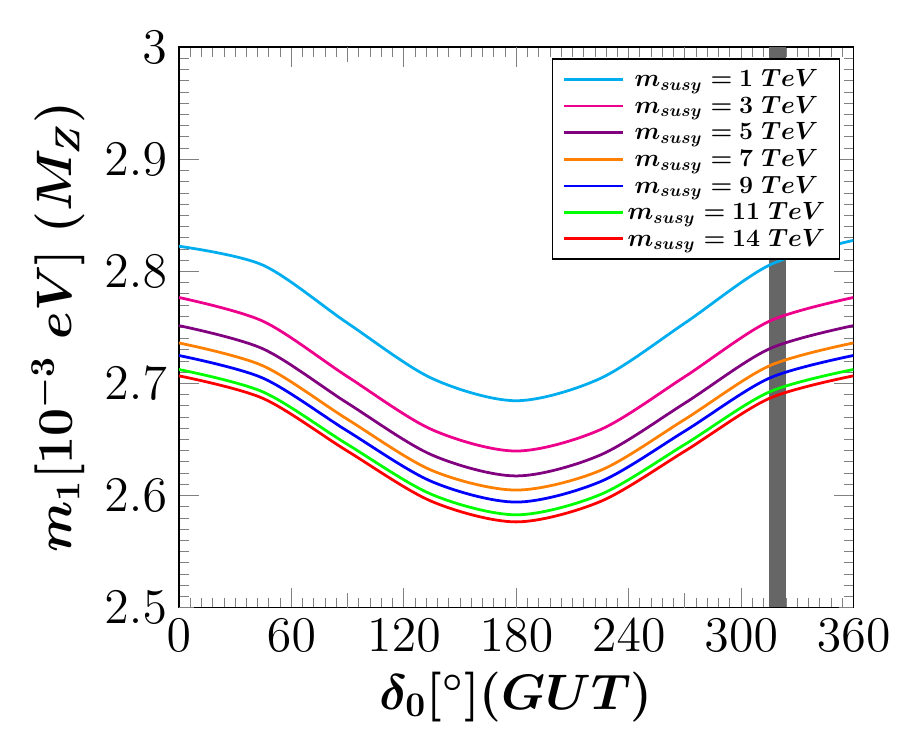} 
\caption{\label{m1}}
\end{subfigure}
\vspace*{\fill}
\begin{subfigure}{0.48\textwidth}
\includegraphics[scale=0.8]{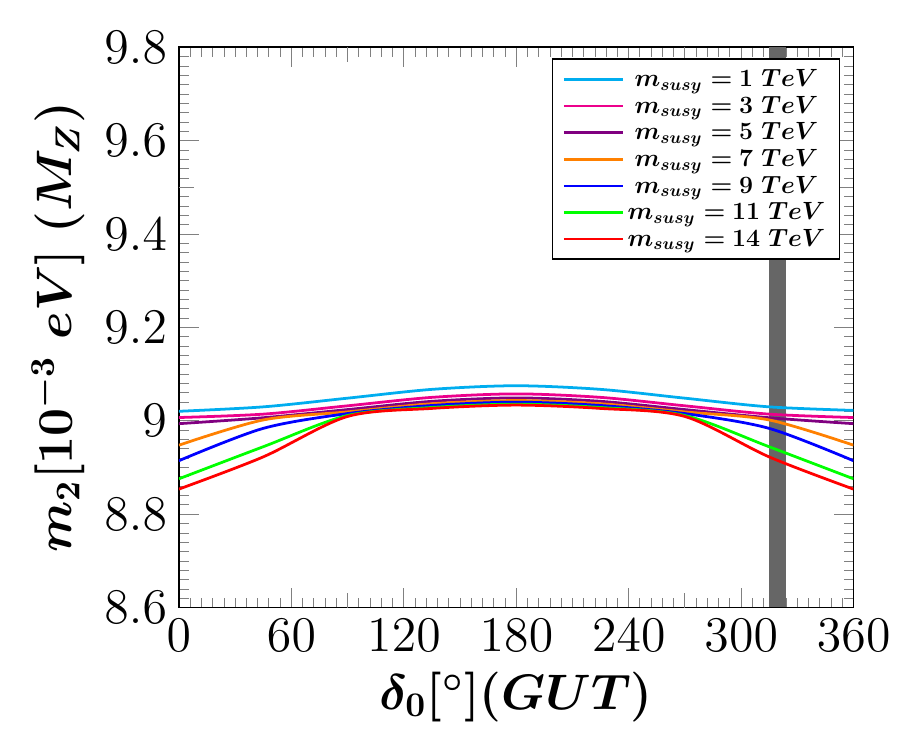} 
\caption{\label{m2}}
\end{subfigure}\\
\begin{subfigure}{0.48\textwidth}
\includegraphics[scale=0.8]{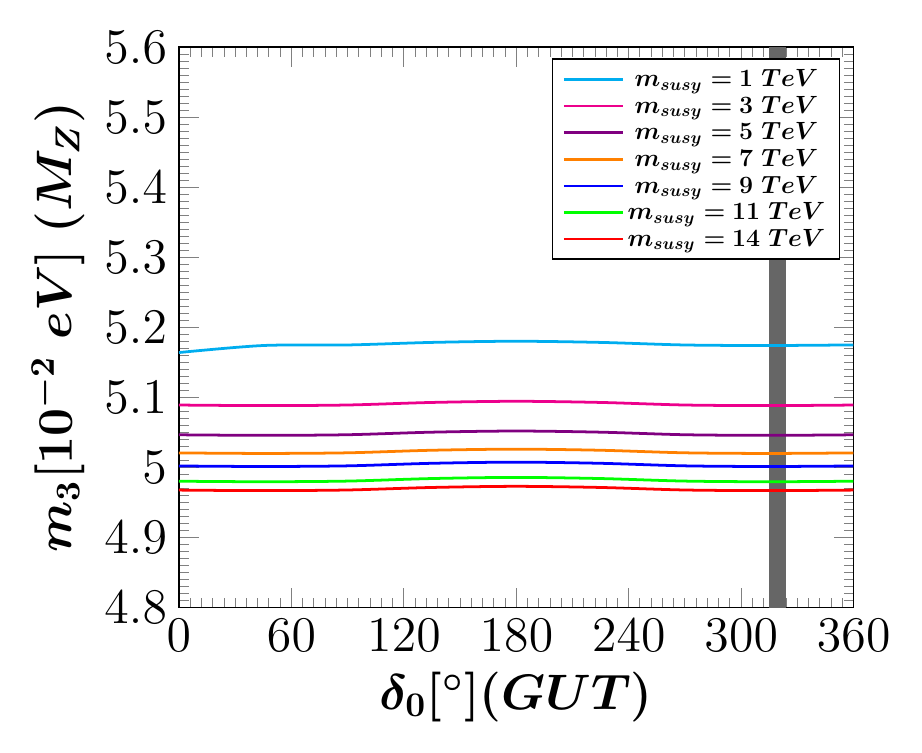} 
\caption{\label{m3}}
\end{subfigure}
\vspace*{\fill}
\begin{subfigure}{0.48\textwidth}
\includegraphics[scale=0.8]{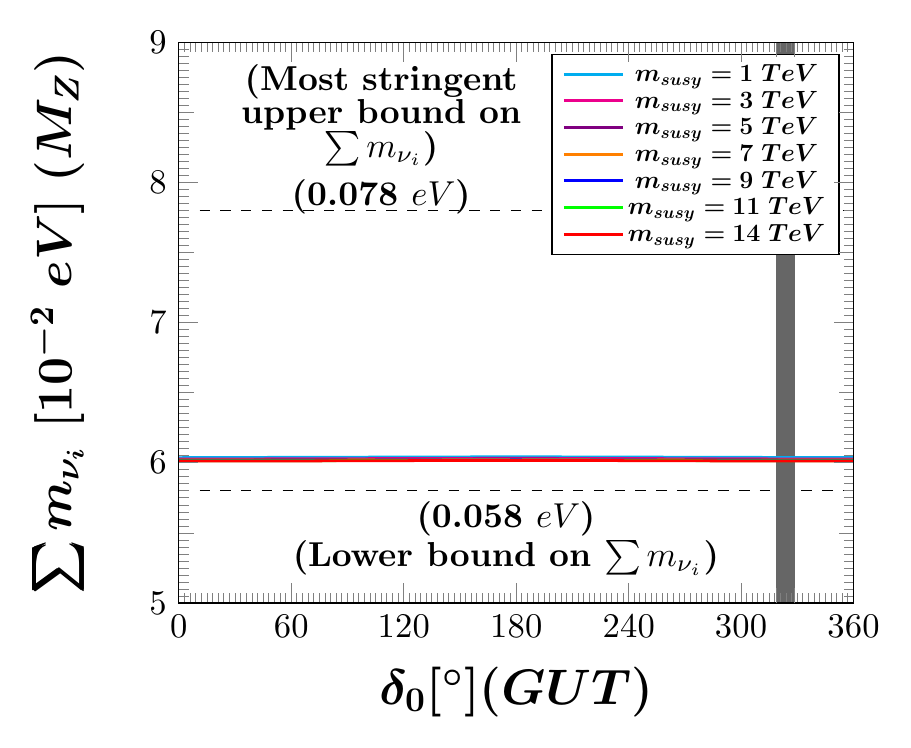} 
\caption{\label{sum}}
\end{subfigure}
\end{center}
\caption{\footnotesize \textbf{(\ref{m1})}, \textbf{(\ref{m2})}, \textbf{(\ref{m3})} and \textbf{(\ref{sum})} show the variation of $m_1$, $m_2$, $m_{3}$ and $\sum\,m_{\nu_i}$ at the $M_{Z}$ scale respectively with respect to the variation of $\delta_{0}$ at the GUT scale for different values of the SUSY breaking scale,\,$m_{susy}$ ranging from $1\,TeV$ to $14\,TeV$. The vertical grey band represents the allowed bound of $\delta_{0}$ at GUT scale which is $318.90^{\circ}\leqslant\delta_{0} \leqslant 328.70^{\circ}$. In Fig.\,(\textbf{\ref{sum}}), the bound on $\sum\,m_{\nu_i}$ is prescribed with respect to the ref.\,\cite{Choudhury:2018byy}.}
\end{figure*}

As the observable $\theta_{13}$ at $M_{z}$ varies a lot with respect to the unphysical parameter $\delta_{0}$, we restrict the latter\,(See Fig.(\ref{t13})) with respect to the $3\sigma$ bound of the former,\cite{deSalas:2017kay}. We see either, $33.40^{\circ}\leqslant\delta_{0}\leqslant 43.30^{\circ}$ or $318.90^{\circ}\leqslant\delta_{0} \leqslant 328.70^{\circ}$. But the first bound predicts a numerical range of the Dirac CP violating phase, $\delta$ at $M_Z$ which lies outside  the $3\sigma$ region and hence it is rejected (See Fig.\,(\ref{dir})). The other bound of $\delta_0$ predicts, $276^{\circ}\leqslant\delta(M_{Z})\leqslant 296^{\circ}$ and this is true with respect to the $2\sigma$ range\,\cite{deSalas:2017kay}.  Now, in view of this allowed range of $\delta_{0}$, one finds, the mixing angles $\theta_{12}$ and $\theta_{23}$ at $M_Z$ scale lie within the $1\sigma$ bound (See Figs.\,(\ref{t12}) and (\ref{t23})). It is found that the mixing angles are less sensitive towards $m_{susy}$. On the contrary, the mass parameters and hence the related observational parameters drift appreciably if $m_{s}$ is varied\,(See Figs.\,(\ref{m1}), (\ref{m2}), (\ref{m3}), (\ref{msol}) and (\ref{matm})). We see that although $\Delta m_{sol}^2$ changes as $m_{susy}$ varies from $1\,TeV$ to $14\,TeV$, yet it agrees well within the $3\sigma$ range. In contrast, the same for $\Delta m_{atm}^2$ goes outside the $3\sigma$ range if $m_{s}\geqslant 9\,TeV$. The $\sum {m_{\nu_{i}}}$ (at $M_{Z}$), varies least with respect to $m_{s}$ and stays within the experimental bound\,(see Fig.\,(\ref{sum})).

We wish to add a few notes on quark masses and mixing parameters obtained at the $M_z$ scale. With the same input parameters at the GUT scale as described towards the beginning of this section along with $m_{susy}$ fixed at $3\,TeV$, it is found that: $m_d (M_z)= 2.349\,MeV$, $m_s(M_z)= 45.092\,MeV$, $m_b(M_z)= 3.019\,GeV$, $m_u(M_z)= 1.254\,MeV$, $m_c(M_z)\approx 0.6141\,GeV$ and $m_t(M_z)= 172.081\,GeV$. These results agree well with the bounds predicted in Ref.\,(\cite{Xing:2011aa}). In addition, we evaluate the CKM mixing parameters at the $M_{Z}$ scale: $|V_{ud}|=0.97443$, $|V_{us}|= 0.22469$,  $|V_{cs}|=0.973616$, $|V_{cb}|=0.040864$, $|V_{ts}|=0.039819$, and $|V_{tb}|=0.99917$. We see $|V_{cb}|$, lies within the $1\sigma$ range, the $|V_{ts}|$ and $|V_{tb}|$ lie within the $2\sigma$ range, whereas the $|V_{ud}|$, $|V_{us}|$ and $|V_{cs}|$ lie within the $3\sigma$ bound\,\cite{Charles:2004jd}. The $|V_{cd}|$ is found to lie slightly below the $3\sigma$ lower limit($0.22452$). We see that the numerical values of $|V_{ub}|\sim \mathcal{O}(10^{-7})$ and $|V_{td}|= 0.00918$. But in reality, $|V_{ub}|\sim 0.003683$ and $|V_{td}|= 0.0085$\,\cite{Charles:2004jd}. The aforesaid inconsistency occurs owing to the fact that while formulating the textures of $Y_d$ and $Y_u$ at GUT scale (see Eqs.\,(\ref{yd}) and (\ref{yu})), the $1$-$3$ rotation within the $V_{CKM}$ is not taken into consideration. We believe that imparting a little perturbation to the textures of $Y_d$ and $Y_u$ may lead to the  necessary changes and further precision to the CKM mixing parameters. This is beyond the scope of the present work. On the other hand we obtain the charged lepton masses as, $m_{e}(M_{Z})\approx 0.4866\,eV $, $m_{\mu}(M_{Z})\approx 102.718\,MeV$ and $m_{\tau}(M_{Z})\approx 1746.09\,MeV$ which are found in agreement with the Ref.\,(\cite{Xing:2011aa}).

\section{Summary}

The present work tries to address the problem of neutrino masses and mixing by looking into the simple unification possibilities and testing the same against the experimental results. We have shown that by relating the smallness of the reactor angle with the Cabibbo angle, unifying both of them at the GUT scale and an extension of the \textit{ansatze} with a Cabibbo motivated GST relation: $\theta_{13}^{\nu}=\theta_{C}=\sqrt{m_{1}/m_{3}}$ for neutrinos as a signature of unification, have got far reaching consequences. Also, this framework considers the unification of $\theta_{12}^{\nu}$ and $\theta_{23}^{\nu}$. Based on the SU(5) GUT phenomenology, we suggest the textures of $Y_{d}$ and $Y_{l}$ such that $\theta_{12}^{l}=1.03\,\theta_{C}$. We run the neutrino mass matrix from $M_{GUT}$ scale to $M_{Z}$ scale following the RGE and explore the oscillation parameters at the the $M_{Z}$ scale. The GST relation within the lepton sector ensures the normal ordering of the neutrino masses. At the $M_{Z}$ scale, the Dirac CP phase is predicted to lie within $276^{\circ}\leqslant\delta(M_{Z})\leqslant 296^{\circ}$ and the $\theta_{23}$ lies in the second octant. We see that the mixing angles are sensitive to the free parameter $\delta_{0}$ whereas, the mass parameters response substantially towards the variation of super symmetry breaking scale.

\section*{Acknowledgment}

SR and KSS thank N. Nimai Singh, Manipur University for the useful discussions. JB thanks Gauhati University for providing him a chance to be a part of the work. 
SR wishes to thank FIST(DST) grant SR/FST/PSI-213/2016(C) for the necessary support.

\bibliographystyle{unsrt}

\begin{thebibliography}{10}
\expandafter\ifx\csname url\endcsname\relax
  \def\url#1{\texttt{#1}}\fi
\expandafter\ifx\csname urlprefix\endcsname\relax\def\urlprefix{URL }\fi
\expandafter\ifx\csname href\endcsname\relax
  \def\href#1#2{#2} \def\path#1{#1}\fi

\bibitem{Pontecorvo:1957cp}
B.~Pontecorvo, {Mesonium and anti-mesonium}, Sov. Phys. JETP 6 (1957) 429, [Zh.
  Eksp. Teor. Fiz.33,549(1957)].

\bibitem{PhysRevD.98.030001}
M.~Tanabashi, et~al., {Review of Particle Physics}, Phys. Rev. D98~(3) (2018)
  030001.
\newblock \href {http://dx.doi.org/10.1103/PhysRevD.98.030001}
  {\path{doi:10.1103/PhysRevD.98.030001}}.

\bibitem{deSalas:2017kay}
P.~F. de~Salas, D.~V. Forero, C.~A. Ternes, M.~Tortola, J.~W.~F. Valle, {Status
  of neutrino oscillations 2018: 3$\sigma$ hint for normal mass ordering and
  improved CP sensitivity}, Phys. Lett. B782 (2018) 633--640.
\newblock \href {http://arxiv.org/abs/1708.01186} {\path{arXiv:1708.01186}},
  \href {http://dx.doi.org/10.1016/j.physletb.2018.06.019}
  {\path{doi:10.1016/j.physletb.2018.06.019}}.

\bibitem{Harrison:2002er}
P.~F. Harrison, D.~H. Perkins, W.~G. Scott, {Tri-bimaximal mixing and the
  neutrino oscillation data}, Phys. Lett. B530 (2002) 167.
\newblock \href {http://arxiv.org/abs/hep-ph/0202074}
  {\path{arXiv:hep-ph/0202074}}, \href
  {http://dx.doi.org/10.1016/S0370-2693(02)01336-9}
  {\path{doi:10.1016/S0370-2693(02)01336-9}}.

\bibitem{An:2012eh}
F.~P. An, et~al., {Observation of electron-antineutrino disappearance at Daya
  Bay}, Phys. Rev. Lett. 108 (2012) 171803.
\newblock \href {http://arxiv.org/abs/1203.1669} {\path{arXiv:1203.1669}},
  \href {http://dx.doi.org/10.1103/PhysRevLett.108.171803}
  {\path{doi:10.1103/PhysRevLett.108.171803}}.

\bibitem{An:2016ses}
F.~P. An, et~al., {Measurement of electron antineutrino oscillation based on
  1230 days of operation of the Daya Bay experiment}, Phys. Rev. D95~(7) (2017)
  072006.
\newblock \href {http://arxiv.org/abs/1610.04802} {\path{arXiv:1610.04802}},
  \href {http://dx.doi.org/10.1103/PhysRevD.95.072006}
  {\path{doi:10.1103/PhysRevD.95.072006}}.

\bibitem{Cabibbo:1977nk}
N.~Cabibbo, {Time Reversal Violation in Neutrino Oscillation}, Phys. Lett. 72B
  (1978) 333--335.
\newblock \href {http://dx.doi.org/10.1016/0370-2693(78)90132-6}
  {\path{doi:10.1016/0370-2693(78)90132-6}}.

\bibitem{Xing:2002sw}
Z.-z. Xing, {Nearly tri bimaximal neutrino mixing and CP violation}, Phys.
  Lett. B 533 (2002) 85--93.
\newblock \href {http://arxiv.org/abs/hep-ph/0204049}
  {\path{arXiv:hep-ph/0204049}}, \href
  {http://dx.doi.org/10.1016/S0370-2693(02)01649-0}
  {\path{doi:10.1016/S0370-2693(02)01649-0}}.

\bibitem{Xing:2011at}
Z.-Z. Xing, {The T2K Indication of Relatively Large $\theta_{13}$ and a Natural
  Perturbation to the Democratic Neutrino Mixing Pattern}, Chin. Phys. C 36
  (2012) 101--105.
\newblock \href {http://arxiv.org/abs/1106.3244} {\path{arXiv:1106.3244}},
  \href {http://dx.doi.org/10.1088/1674-1137/36/2/001}
  {\path{doi:10.1088/1674-1137/36/2/001}}.

\bibitem{Boucenna:2012xb}
S.~M. Boucenna, S.~Morisi, M.~Tortola, J.~W.~F. Valle, {Bi-large neutrino
  mixing and the Cabibbo angle}, Phys. Rev. D86 (2012) 051301.
\newblock \href {http://arxiv.org/abs/1206.2555} {\path{arXiv:1206.2555}},
  \href {http://dx.doi.org/10.1103/PhysRevD.86.051301}
  {\path{doi:10.1103/PhysRevD.86.051301}}.

\bibitem{Ding:2012wh}
G.-J. Ding, S.~Morisi, J.~W.~F. Valle, {Bilarge neutrino mixing and Abelian
  flavor symmetry}, Phys. Rev. D87~(5) (2013) 053013.
\newblock \href {http://arxiv.org/abs/1211.6506} {\path{arXiv:1211.6506}},
  \href {http://dx.doi.org/10.1103/PhysRevD.87.053013}
  {\path{doi:10.1103/PhysRevD.87.053013}}.

\bibitem{Branco:2014zza}
G.~C. Branco, M.~N. Rebelo, J.~I. Silva-Marcos, D.~Wegman, {Quasidegeneracy of
  Majorana Neutrinos and the Origin of Large Leptonic Mixing}, Phys. Rev.
  D91~(1) (2015) 013001.
\newblock \href {http://arxiv.org/abs/1405.5120} {\path{arXiv:1405.5120}},
  \href {http://dx.doi.org/10.1103/PhysRevD.91.013001}
  {\path{doi:10.1103/PhysRevD.91.013001}}.

\bibitem{Roy:2012ib}
S.~Roy, N.~N. Singh, {Bi-Large neutrino mixing with charged lepton correction},
  Indian J. Phys. 88~(5) (2014) 513--519.
\newblock \href {http://arxiv.org/abs/1211.7207} {\path{arXiv:1211.7207}},
  \href {http://dx.doi.org/10.1007/s12648-014-0446-1}
  {\path{doi:10.1007/s12648-014-0446-1}}.

\bibitem{Roy:2014nua}
S.~Roy, S.~Morisi, N.~N. Singh, J.~W.~F. Valle, {The Cabibbo angle as a
  universal seed for quark and lepton mixings}, Phys. Lett. B748 (2015) 1--4.
\newblock \href {http://arxiv.org/abs/1410.3658} {\path{arXiv:1410.3658}},
  \href {http://dx.doi.org/10.1016/j.physletb.2015.06.052}
  {\path{doi:10.1016/j.physletb.2015.06.052}}.

\bibitem{Ding:2019vvi}
G.-J. Ding, N.~Nath, R.~Srivastava, J.~W.~F. Valle, {Status and prospects of
  ‘bi-large’ leptonic mixing}, Phys. Lett. B796 (2019) 162--167.
\newblock \href {http://arxiv.org/abs/1904.05632} {\path{arXiv:1904.05632}},
  \href {http://dx.doi.org/10.1016/j.physletb.2019.07.037}
  {\path{doi:10.1016/j.physletb.2019.07.037}}.

\bibitem{Chen:2019egu}
P.~Chen, G.-J. Ding, R.~Srivastava, J.~W.~F. Valle, {Predicting neutrino
  oscillations with “bi-large” lepton mixing matrices}, Phys. Lett. B792
  (2019) 461--464.
\newblock \href {http://arxiv.org/abs/1902.08962} {\path{arXiv:1902.08962}},
  \href {http://dx.doi.org/10.1016/j.physletb.2019.04.022}
  {\path{doi:10.1016/j.physletb.2019.04.022}}.

\bibitem{Wolfenstein:1983yz}
L.~Wolfenstein, {Parametrization of the Kobayashi-Maskawa Matrix}, Phys. Rev.
  Lett. 51 (1983) 1945.
\newblock \href {http://dx.doi.org/10.1103/PhysRevLett.51.1945}
  {\path{doi:10.1103/PhysRevLett.51.1945}}.

\bibitem{Pati:1974yy}
J.~C. Pati, A.~Salam, {Lepton Number as the Fourth Color}, Phys. Rev. D10
  (1974) 275--289, [Erratum: Phys. Rev.D11,703(1975)].
\newblock \href {http://dx.doi.org/10.1103/PhysRevD.10.275,
  10.1103/PhysRevD.11.703.2} {\path{doi:10.1103/PhysRevD.10.275,
  10.1103/PhysRevD.11.703.2}}.

\bibitem{Elias:1975kf}
V.~Elias, A.~R. Swift, {Generalization of the Pati-Salam Model}, Phys. Rev. D13
  (1976) 2083.
\newblock \href {http://dx.doi.org/10.1103/PhysRevD.13.2083}
  {\path{doi:10.1103/PhysRevD.13.2083}}.

\bibitem{Elias:1977bv}
V.~Elias, {Gauge Coupling Constant Magnitudes in the Pati-Salam Model}, Phys.
  Rev. D16 (1977) 1586.
\newblock \href {http://dx.doi.org/10.1103/PhysRevD.16.1586}
  {\path{doi:10.1103/PhysRevD.16.1586}}.

\bibitem{Blazek:2003wz}
T.~Blazek, S.~F. King, J.~K. Parry, {Global analysis of a supersymmetric
  Pati-Salam model}, JHEP 05 (2003) 016.
\newblock \href {http://arxiv.org/abs/hep-ph/0303192}
  {\path{arXiv:hep-ph/0303192}}, \href
  {http://dx.doi.org/10.1088/1126-6708/2003/05/016}
  {\path{doi:10.1088/1126-6708/2003/05/016}}.

\bibitem{Dent:2007eu}
J.~B. Dent, T.~W. Kephart, {Minimal Pati-Salam model from string theory
  unification}, Phys. Rev. D77 (2008) 115008.
\newblock \href {http://arxiv.org/abs/0705.1995} {\path{arXiv:0705.1995}},
  \href {http://dx.doi.org/10.1103/PhysRevD.77.115008}
  {\path{doi:10.1103/PhysRevD.77.115008}}.

\bibitem{kounnas1985grand}
C.~Kounnas, A.~Masiero, D.~Nanopoulos, K.~Olive,
  \href{https://books.google.co.in/books?id=XAc8DQAAQBAJ}{Grand Unification
  with and without Supersymmetry and Cosmological Implications}, International
  School for Advanced Studies Lecture Series, Singapore, 1985.
\newline\urlprefix\url{https://books.google.co.in/books?id=XAc8DQAAQBAJ}

\bibitem{ross2003grand}
G.~Ross, \href{https://books.google.co.in/books?id=ccj\_swEACAAJ}{Grand Unified
  Theories}, Frontiers in Physics, Avalon Publishing, 2003.
\newline\urlprefix\url{https://books.google.co.in/books?id=ccj\_swEACAAJ}

\bibitem{10.1143/PTP.49.652}
M.~Kobayashi, T.~Maskawa,
  \href{https://doi.org/10.1143/PTP.49.652}{{CP-Violation in the Renormalizable
  Theory of Weak Interaction}}, Progress of Theoretical Physics 49~(2) (1973)
  652--657.
\newblock \href
  {http://arxiv.org/abs/http://oup.prod.sis.lan/ptp/article-pdf/49/2/652/5257692/49-2-652.pdf}
  {\path{arXiv:http://oup.prod.sis.lan/ptp/article-pdf/49/2/652/5257692/49-2-652.pdf}},
  \href {http://dx.doi.org/10.1143/PTP.49.652} {\path{doi:10.1143/PTP.49.652}}.
\newline\urlprefix\url{https://doi.org/10.1143/PTP.49.652}

\bibitem{Duarah:2012bd}
C.~Duarah, A.~Das, N.~N. Singh, {Charged lepton contributions to bimaximal and
  tri-bimaximal mixings for generating $\sin\theta_{13}\neq 0$ and
  $\tan^2\theta_{23}<1$}, Phys. Lett. B718 (2012) 147--152.
\newblock \href {http://arxiv.org/abs/1207.5225} {\path{arXiv:1207.5225}},
  \href {http://dx.doi.org/10.1016/j.physletb.2012.10.033}
  {\path{doi:10.1016/j.physletb.2012.10.033}}.

\bibitem{XING20201}
Z.~zhong Xing,
  \href{http://www.sciencedirect.com/science/article/pii/S0370157320300223}{Flavor
  structures of charged fermions and massive neutrinos}, Physics Reports 854
  (2020) 1 -- 147, flavor structures of charged fermions and massive neutrinos.
\newblock \href
  {http://dx.doi.org/https://doi.org/10.1016/j.physrep.2020.02.001}
  {\path{doi:https://doi.org/10.1016/j.physrep.2020.02.001}}.
\newline\urlprefix\url{http://www.sciencedirect.com/science/article/pii/S0370157320300223}

\bibitem{Gatto:1968ss}
R.~Gatto, G.~Sartori, M.~Tonin, {Weak Selfmasses, Cabibbo Angle, and Broken
  SU(2) x SU(2)}, Phys. Lett. 28B (1968) 128--130.
\newblock \href {http://dx.doi.org/10.1016/0370-2693(68)90150-0}
  {\path{doi:10.1016/0370-2693(68)90150-0}}.

\bibitem{doi:10.1111/j.2164-0947.1977.tb02958.x}
S.~Weinberg,
  \href{https://nyaspubs.onlinelibrary.wiley.com/doi/abs/10.1111/j.2164-0947.1977.tb02958.x}{The
  problem of mass}, Transactions of the New York Academy of Sciences 38~(1
  Series II) (1977) 185--201.
\newblock \href
  {http://arxiv.org/abs/https://nyaspubs.onlinelibrary.wiley.com/doi/pdf/10.1111/j.2164-0947.1977.tb02958.x}
  {\path{arXiv:https://nyaspubs.onlinelibrary.wiley.com/doi/pdf/10.1111/j.2164-0947.1977.tb02958.x}},
  \href {http://dx.doi.org/10.1111/j.2164-0947.1977.tb02958.x}
  {\path{doi:10.1111/j.2164-0947.1977.tb02958.x}}.
\newline\urlprefix\url{https://nyaspubs.onlinelibrary.wiley.com/doi/abs/10.1111/j.2164-0947.1977.tb02958.x}

\bibitem{WILCZEK1977418}
F.~Wilczek, Z.~Zee,
  \href{http://www.sciencedirect.com/science/article/pii/0370269377904038}{Discrete
  flavor symmetries and a formula for the cabibbo angle}, Physics Letters B
  70~(4) (1977) 418 -- 420.
\newblock \href
  {http://dx.doi.org/https://doi.org/10.1016/0370-2693(77)90403-8}
  {\path{doi:https://doi.org/10.1016/0370-2693(77)90403-8}}.
\newline\urlprefix\url{http://www.sciencedirect.com/science/article/pii/0370269377904038}

\bibitem{FRITZSCH1977436}
H.~Fritzsch,
  \href{http://www.sciencedirect.com/science/article/pii/0370269377904087}{Calculating
  the cabibbo angle}, Physics Letters B 70~(4) (1977) 436 -- 440.
\newblock \href
  {http://dx.doi.org/https://doi.org/10.1016/0370-2693(77)90408-7}
  {\path{doi:https://doi.org/10.1016/0370-2693(77)90408-7}}.
\newline\urlprefix\url{http://www.sciencedirect.com/science/article/pii/0370269377904087}

\bibitem{PAKVASA197861}
S.~Pakvasa, H.~Sugawara,
  \href{http://www.sciencedirect.com/science/article/pii/0370269378901727}{Discrete
  symmetry and cabibbo angle}, Physics Letters B 73~(1) (1978) 61 -- 64.
\newblock \href
  {http://dx.doi.org/https://doi.org/10.1016/0370-2693(78)90172-7}
  {\path{doi:https://doi.org/10.1016/0370-2693(78)90172-7}}.
\newline\urlprefix\url{http://www.sciencedirect.com/science/article/pii/0370269378901727}

\bibitem{Roy:2015cza}
S.~Roy, N.~N. Singh, {Mixing angle as a function of neutrino mass ratio}, Phys.
  Rev. D91~(9) (2015) 096003.
\newblock \href {http://arxiv.org/abs/1603.07474} {\path{arXiv:1603.07474}},
  \href {http://dx.doi.org/10.1103/PhysRevD.91.096003}
  {\path{doi:10.1103/PhysRevD.91.096003}}.

\bibitem{Charles:2004jd}
J.~Charles, A.~Hocker, H.~Lacker, S.~Laplace, F.~Le~Diberder, J.~Malcles,
  J.~Ocariz, M.~Pivk, L.~Roos, {CP violation and the CKM matrix: Assessing the
  impact of the asymmetric $B$ factories}, Eur. Phys. J. C 41~(1) (2005)
  1--131.
\newblock \href {http://arxiv.org/abs/hep-ph/0406184}
  {\path{arXiv:hep-ph/0406184}}, \href
  {http://dx.doi.org/10.1140/epjc/s2005-02169-1}
  {\path{doi:10.1140/epjc/s2005-02169-1}}.

\bibitem{Minkowski:1977sc}
P.~Minkowski, {$\mu \to e\gamma$ at a Rate of One Out of $10^{9}$ Muon
  Decays?}, Phys. Lett. 67B (1977) 421--428.
\newblock \href {http://dx.doi.org/10.1016/0370-2693(77)90435-X}
  {\path{doi:10.1016/0370-2693(77)90435-X}}.

\bibitem{Mohapatra:1979ia}
R.~N. Mohapatra, G.~Senjanovic, {Neutrino Mass and Spontaneous Parity
  Nonconservation}, Phys. Rev. Lett. 44 (1980) 912, [,231(1979)].
\newblock \href {http://dx.doi.org/10.1103/PhysRevLett.44.912}
  {\path{doi:10.1103/PhysRevLett.44.912}}.

\bibitem{Randall:1998uk}
L.~Randall, R.~Sundrum, {Out of this world supersymmetry breaking}, Nucl. Phys.
  B557 (1999) 79--118.
\newblock \href {http://arxiv.org/abs/hep-th/9810155}
  {\path{arXiv:hep-th/9810155}}, \href
  {http://dx.doi.org/10.1016/S0550-3213(99)00359-4}
  {\path{doi:10.1016/S0550-3213(99)00359-4}}.

\bibitem{Haber:1990aw}
H.~E. Haber, R.~Hempfling, {Can the mass of the lightest Higgs boson of the
  minimal supersymmetric model be larger than m(Z)?}, Phys. Rev. Lett. 66
  (1991) 1815--1818.
\newblock \href {http://dx.doi.org/10.1103/PhysRevLett.66.1815}
  {\path{doi:10.1103/PhysRevLett.66.1815}}.

\bibitem{Dine:1995ag}
M.~Dine, A.~E. Nelson, Y.~Nir, Y.~Shirman, {New tools for low-energy dynamical
  supersymmetry breaking}, Phys. Rev. D53 (1996) 2658--2669.
\newblock \href {http://arxiv.org/abs/hep-ph/9507378}
  {\path{arXiv:hep-ph/9507378}}, \href
  {http://dx.doi.org/10.1103/PhysRevD.53.2658}
  {\path{doi:10.1103/PhysRevD.53.2658}}.

\bibitem{Balaji:2000au}
K.~R.~S. Balaji, A.~S. Dighe, R.~N. Mohapatra, M.~K. Parida, {Radiative
  magnification of neutrino mixings and a natural explanation of the neutrino
  anomalies}, Phys. Lett. B481 (2000) 33--38.
\newblock \href {http://arxiv.org/abs/hep-ph/0002177}
  {\path{arXiv:hep-ph/0002177}}, \href
  {http://dx.doi.org/10.1016/S0370-2693(00)00410-X}
  {\path{doi:10.1016/S0370-2693(00)00410-X}}.

\bibitem{Mohapatra:2003tw}
R.~N. Mohapatra, M.~K. Parida, G.~Rajasekaran, {High scale mixing unification
  and large neutrino mixing angles}, Phys. Rev. D69 (2004) 053007.
\newblock \href {http://arxiv.org/abs/hep-ph/0301234}
  {\path{arXiv:hep-ph/0301234}}, \href
  {http://dx.doi.org/10.1103/PhysRevD.69.053007}
  {\path{doi:10.1103/PhysRevD.69.053007}}.

\bibitem{Antusch:2002fr}
S.~Antusch, M.~Ratz, {Radiative generation of the LMA solution from small solar
  neutrino mixing at the GUT scale}, JHEP 11 (2002) 010.
\newblock \href {http://arxiv.org/abs/hep-ph/0208136}
  {\path{arXiv:hep-ph/0208136}}, \href
  {http://dx.doi.org/10.1088/1126-6708/2002/11/010}
  {\path{doi:10.1088/1126-6708/2002/11/010}}.

\bibitem{Joshipura:2002xa}
A.~S. Joshipura, S.~D. Rindani, N.~N. Singh, {Predictive framework with a pair
  of degenerate neutrinos at a high scale}, Nucl. Phys. B660 (2003) 362--372.
\newblock \href {http://arxiv.org/abs/hep-ph/0211378}
  {\path{arXiv:hep-ph/0211378}}, \href
  {http://dx.doi.org/10.1016/S0550-3213(03)00236-0}
  {\path{doi:10.1016/S0550-3213(03)00236-0}}.

\bibitem{Antusch:2003kp}
S.~Antusch, J.~Kersten, M.~Lindner, M.~Ratz, {Running neutrino masses, mixings
  and CP phases: Analytical results and phenomenological consequences}, Nucl.
  Phys. B674 (2003) 401--433.
\newblock \href {http://arxiv.org/abs/hep-ph/0305273}
  {\path{arXiv:hep-ph/0305273}}, \href
  {http://dx.doi.org/10.1016/j.nuclphysb.2003.09.050}
  {\path{doi:10.1016/j.nuclphysb.2003.09.050}}.

\bibitem{Ohlsson:2013xva}
T.~Ohlsson, S.~Zhou, {Renormalization group running of neutrino parameters},
  Nature Commun. 5 (2014) 5153.
\newblock \href {http://arxiv.org/abs/1311.3846} {\path{arXiv:1311.3846}},
  \href {http://dx.doi.org/10.1038/ncomms6153} {\path{doi:10.1038/ncomms6153}}.

\bibitem{Gupta:2014lwa}
S.~Gupta, S.~K. Kang, C.~S. Kim, {Renormalization Group Evolution of Neutrino
  Parameters in Presence of Seesaw Threshold Effects and Majorana Phases},
  Nucl. Phys. B893 (2015) 89--106.
\newblock \href {http://arxiv.org/abs/1406.7476} {\path{arXiv:1406.7476}},
  \href {http://dx.doi.org/10.1016/j.nuclphysb.2015.01.026}
  {\path{doi:10.1016/j.nuclphysb.2015.01.026}}.

\bibitem{Haba:2014uza}
N.~Haba, K.~Kaneta, R.~Takahashi, Y.~Yamaguchi, {Accurate renormalization group
  analyses in neutrino sector}, Nucl. Phys. B885 (2014) 180--195.
\newblock \href {http://arxiv.org/abs/1402.4126} {\path{arXiv:1402.4126}},
  \href {http://dx.doi.org/10.1016/j.nuclphysb.2014.05.022}
  {\path{doi:10.1016/j.nuclphysb.2014.05.022}}.

\bibitem{Hollik:2014hya}
W.~G. Hollik, {Radiative generation of neutrino mixing: degenerate masses and
  threshold corrections}, Phys. Rev. D91~(3) (2015) 033001.
\newblock \href {http://arxiv.org/abs/1412.4585} {\path{arXiv:1412.4585}},
  \href {http://dx.doi.org/10.1103/PhysRevD.91.033001}
  {\path{doi:10.1103/PhysRevD.91.033001}}.

\bibitem{Singh:2015kua}
K.~S. Singh, N.~N. Singh, {Effects of the Variation of SUSY Breaking Scale on
  Yukawa and Gauge Couplings Unification}, Adv. High Energy Phys. 2015 (2015)
  652029.
\newblock \href {http://dx.doi.org/10.1155/2015/652029}
  {\path{doi:10.1155/2015/652029}}.

\bibitem{Singh:2018cxy}
K.~S. Singh, S.~Roy, N.~N. Singh, {Stability of neutrino parameters and
  self-complementarity relation with varying SUSY breaking scale}, Phys. Rev.
  D97~(5) (2018) 055038.
\newblock \href {http://arxiv.org/abs/1802.09784} {\path{arXiv:1802.09784}},
  \href {http://dx.doi.org/10.1103/PhysRevD.97.055038}
  {\path{doi:10.1103/PhysRevD.97.055038}}.

\bibitem{Antusch:2009gu}
S.~Antusch, M.~Spinrath, {New GUT predictions for quark and lepton mass ratios
  confronted with phenomenology}, Phys. Rev. D79 (2009) 095004.
\newblock \href {http://arxiv.org/abs/0902.4644} {\path{arXiv:0902.4644}},
  \href {http://dx.doi.org/10.1103/PhysRevD.79.095004}
  {\path{doi:10.1103/PhysRevD.79.095004}}.

\bibitem{Marzocca:2011dh}
D.~Marzocca, S.~T. Petcov, A.~Romanino, M.~Spinrath, {Sizeable $\theta_{13}$
  from the Charged Lepton Sector in SU(5), (Tri-)Bimaximal Neutrino Mixing and
  Dirac CP Violation}, JHEP 11 (2011) 009.
\newblock \href {http://arxiv.org/abs/1108.0614} {\path{arXiv:1108.0614}},
  \href {http://dx.doi.org/10.1007/JHEP11(2011)009}
  {\path{doi:10.1007/JHEP11(2011)009}}.

\bibitem{Antusch:2013ti}
S.~Antusch, {Models for Neutrino Masses and Mixings}[Nucl. Phys. Proc.
  Suppl.235-236,303(2013)].
\newblock \href {http://arxiv.org/abs/1301.5511} {\path{arXiv:1301.5511}},
  \href {http://dx.doi.org/10.1016/j.nuclphysbps.2013.04.026}
  {\path{doi:10.1016/j.nuclphysbps.2013.04.026}}.

\bibitem{Antusch:2013rxa}
S.~Antusch, S.~F. King, M.~Spinrath, {GUT predictions for quark-lepton Yukawa
  coupling ratios with messenger masses from non-singlets}, Phys. Rev. D89~(5)
  (2014) 055027.
\newblock \href {http://arxiv.org/abs/1311.0877} {\path{arXiv:1311.0877}},
  \href {http://dx.doi.org/10.1103/PhysRevD.89.055027}
  {\path{doi:10.1103/PhysRevD.89.055027}}.

\bibitem{Roy:2015cka}
S.~Roy, N.~N. Singh, {Modulated bimaximal neutrino mixing}, Phys. Rev. D92~(3)
  (2015) 036001.
\newblock \href {http://arxiv.org/abs/1603.07972} {\path{arXiv:1603.07972}},
  \href {http://dx.doi.org/10.1103/PhysRevD.92.036001}
  {\path{doi:10.1103/PhysRevD.92.036001}}.

\bibitem{Rahat:2018sgs}
M.~H. Rahat, P.~Ramond, B.~Xu, {Asymmetric tribimaximal texture}, Phys. Rev.
  D98~(5) (2018) 055030.
\newblock \href {http://arxiv.org/abs/1805.10684} {\path{arXiv:1805.10684}},
  \href {http://dx.doi.org/10.1103/PhysRevD.98.055030}
  {\path{doi:10.1103/PhysRevD.98.055030}}.

\bibitem{Antusch:2005gp}
S.~Antusch, J.~Kersten, M.~Lindner, M.~Ratz, M.~A. Schmidt, {Running neutrino
  mass parameters in see-saw scenarios}, JHEP 03 (2005) 024.
\newblock \href {http://arxiv.org/abs/hep-ph/0501272}
  {\path{arXiv:hep-ph/0501272}}, \href
  {http://dx.doi.org/10.1088/1126-6708/2005/03/024}
  {\path{doi:10.1088/1126-6708/2005/03/024}}.

\bibitem{Tanimoto:1995bf}
M.~Tanimoto, {Renormalization effect on large neutrino flavor mixing in the
  minimal supersymmetric standard model}, Phys. Lett. B360 (1995) 41--46.
\newblock \href {http://arxiv.org/abs/hep-ph/9508247}
  {\path{arXiv:hep-ph/9508247}}, \href
  {http://dx.doi.org/10.1016/0370-2693(95)01107-2}
  {\path{doi:10.1016/0370-2693(95)01107-2}}.

\bibitem{Mei:2004rn}
J.-w. Mei, Z.-z. Xing, {Radiative generation of theta(13) with the seesaw
  threshold effect}, Phys. Rev. D70 (2004) 053002.
\newblock \href {http://arxiv.org/abs/hep-ph/0404081}
  {\path{arXiv:hep-ph/0404081}}, \href
  {http://dx.doi.org/10.1103/PhysRevD.70.053002}
  {\path{doi:10.1103/PhysRevD.70.053002}}.

\bibitem{Ohlsson:2019sja}
T.~Ohlsson, M.~Pernow, {Fits to Non-Supersymmetric SO(10) Models with Type I
  and II Seesaw Mechanisms Using Renormalization Group Evolution}, JHEP 06
  (2019) 085.
\newblock \href {http://arxiv.org/abs/1903.08241} {\path{arXiv:1903.08241}},
  \href {http://dx.doi.org/10.1007/JHEP06(2019)085}
  {\path{doi:10.1007/JHEP06(2019)085}}.

\bibitem{Coy:2018bxr}
R.~Coy, M.~Frigerio, {Effective approach to lepton observables: the seesaw
  case}, Phys. Rev. D99~(9) (2019) 095040.
\newblock \href {http://arxiv.org/abs/1812.03165} {\path{arXiv:1812.03165}},
  \href {http://dx.doi.org/10.1103/PhysRevD.99.095040}
  {\path{doi:10.1103/PhysRevD.99.095040}}.

\bibitem{Antusch:2018gnu}
S.~Antusch, C.~Hohl, C.~K. Khosa, V.~Susic, {Predicting $\delta^\text{PMNS}$,
  $\theta_{23}^\text{PMNS}$ and fermion mass ratios from flavour GUTs with
  CSD2}, JHEP 12 (2018) 025.
\newblock \href {http://arxiv.org/abs/1808.09364} {\path{arXiv:1808.09364}},
  \href {http://dx.doi.org/10.1007/JHEP12(2018)025}
  {\path{doi:10.1007/JHEP12(2018)025}}.

\bibitem{Huan:2018lzd}
G.-y. Huang, Z.-z. Xing, J.-y. Zhu, {Correlation of normal neutrino mass
  ordering with upper octant of $\theta^{}_{23}$ and third quadrant of $\delta$
  via RGE-induced $\mu$-$\tau$ symmetry breaking}, Chin. Phys. C42~(12) (2018)
  123108.
\newblock \href {http://arxiv.org/abs/1806.06640} {\path{arXiv:1806.06640}},
  \href {http://dx.doi.org/10.1088/1674-1137/42/12/123108}
  {\path{doi:10.1088/1674-1137/42/12/123108}}.

\bibitem{Antusch:2017ano}
S.~Antusch, C.~Hohl, {Predictions from a flavour GUT model combined with a SUSY
  breaking sector}, JHEP 10 (2017) 155.
\newblock \href {http://arxiv.org/abs/1706.04274} {\path{arXiv:1706.04274}},
  \href {http://dx.doi.org/10.1007/JHEP10(2017)155}
  {\path{doi:10.1007/JHEP10(2017)155}}.

\bibitem{Esteban:2018azc}
I.~Esteban, M.~Gonzalez-Garcia, A.~Hernandez-Cabezudo, M.~Maltoni, T.~Schwetz,
  {Global analysis of three-flavour neutrino oscillations: synergies and
  tensions in the determination of $\theta_{23}$, $\delta_{CP}$, and the mass
  ordering}, JHEP 01 (2019) 106.
\newblock \href {http://arxiv.org/abs/1811.05487} {\path{arXiv:1811.05487}},
  \href {http://dx.doi.org/10.1007/JHEP01(2019)106}
  {\path{doi:10.1007/JHEP01(2019)106}}.

\bibitem{Giusarma:2016phn}
E.~Giusarma, M.~Gerbino, O.~Mena, S.~Vagnozzi, S.~Ho, K.~Freese, {Improvement
  of cosmological neutrino mass bounds}, Phys. Rev. D94~(8) (2016) 083522.
\newblock \href {http://arxiv.org/abs/1605.04320} {\path{arXiv:1605.04320}},
  \href {http://dx.doi.org/10.1103/PhysRevD.94.083522}
  {\path{doi:10.1103/PhysRevD.94.083522}}.

\bibitem{Loureiro:2018pdz}
A.~Loureiro, et~al., {On The Upper Bound of Neutrino Masses from Combined
  Cosmological Observations and Particle Physics Experiments}, Phys. Rev. Lett.
  123~(8) (2019) 081301.
\newblock \href {http://arxiv.org/abs/1811.02578} {\path{arXiv:1811.02578}},
  \href {http://dx.doi.org/10.1103/PhysRevLett.123.081301}
  {\path{doi:10.1103/PhysRevLett.123.081301}}.

\bibitem{Choudhury:2018byy}
S.~Roy~Choudhury, S.~Choubey, {Updated Bounds on Sum of Neutrino Masses in
  Various Cosmological Scenarios}, JCAP 1809~(09) (2018) 017.
\newblock \href {http://arxiv.org/abs/1806.10832} {\path{arXiv:1806.10832}},
  \href {http://dx.doi.org/10.1088/1475-7516/2018/09/017}
  {\path{doi:10.1088/1475-7516/2018/09/017}}.

\bibitem{Xing:2011aa}
Z.-z. Xing, H.~Zhang, S.~Zhou, {Impacts of the Higgs mass on vacuum stability,
  running fermion masses and two-body Higgs decays}, Phys. Rev. D 86 (2012)
  013013.
\newblock \href {http://arxiv.org/abs/1112.3112} {\path{arXiv:1112.3112}},
  \href {http://dx.doi.org/10.1103/PhysRevD.86.013013}
  {\path{doi:10.1103/PhysRevD.86.013013}}.

\end{thebibliography}

\newpage


\begin{table}
\caption{\footnotesize The coefficients of $Y_{d}$ and $Y_{u}$ as shown in eqs.\,(\ref{yd}) and (\ref{yu}) respectively are described in this table.}
\label{Table1}
\begin{tabular}{l}
\hline\noalign{\smallskip}
$\mathcal{O}(1)$ coefficients appearing in $Y_{d,u,\nu}$ \\
\hline\noalign{\smallskip}
 \begin{tabular}{c}
\begin{small}$ d_{12}=2.756,\,\, d_{21}= 2.7118,\,\,d_{22}=2.4984,$\end{small} \\ 
\begin{small}$ d_{23}=1.9119,\,\, d_{31}= 1.9130,\,\, d_{32}=1.7625,\,\, d_{33}=2.7102$\end{small} \\ 
\end{tabular} \\
\hline\noalign{\smallskip}
\begin{small} $u_{11}=0.7718,\,\, u_{22}=0.9836,\,\,u_{23}=0.6938,\,\,u_{32}=1.5168,\,\, u_{33}=0.4827$  \end{small}\\
\hline\noalign{\smallskip}
\begin{small} $\nu_{11}=0.8733,\,\, \nu_{21}=0.7626\,i,\,\,\nu_{22}=0.4437$ \end{small} \\
\hline\noalign{\smallskip}
\end{tabular}
\end{table}

\begin{table}
\caption{ The list of the gauge coupling constants $g_1$, $g_2$ , $g_3$ and the $M_{GUT}$ for different values of the SUSY breaking scales ranging from $1\,TeV$  to $14\,TeV$ is given.}
\label{Table2}
\begin{tabular}{ccccc}
\hline\noalign{\smallskip}
$m_{s}(TeV)$ & $M_{GUT}(10^{16}\,GeV)$ & $g_1$   & $g_2$		  & $g_3$\\
\hline\noalign{\smallskip}
1  	   & 4.090 		&  0.7151 	 & 0.7154  		  & 0.7158\\
\hline\noalign{\smallskip}
3       & 4.577     & 0.7063      &  0.7065        & 0.7069\\
\hline\noalign{\smallskip}
5       & 4.790     & 0.7028      &  0.7031        & 0.7034\\
\hline\noalign{\smallskip}
7       & 4.848     & 0.7007      &  0.7009        & 0.7010\\
\hline\noalign{\smallskip}
9       & 4.912     & 0.6987      &  0.6987        & 0.6987\\
\hline\noalign{\smallskip}       
11      & 5.112     & 0.6973      &  0.6975        & 0.6977\\
\hline\noalign{\smallskip}
14      & 7.211     & 0.6954      &  0.6957        & 0.6916\\
\hline\noalign{\smallskip}
\end{tabular}
\end{table}

\end{document}